\documentclass[apj,numberedappendix]{emulateapj}
\usepackage{xcolor}
\usepackage{amsmath}
\usepackage[colorlinks=true,citecolor=black,linkcolor=black,urlcolor=blue]{hyperref}

\newcommand{\msunh}{\>h^{-1}\rm M_\odot}

\newcommand{\mpch}{\>h^{-1}{\rm {Mpc}}}

\newcommand{\rmd}{{\rm d}}

\newcommand{\kpch}{\>h^{-1}{\rm {kpc}}}

\usepackage{color}
\definecolor{darkgreen}{rgb}{0.0,0.5,0.0}
\usepackage{array}
\usepackage{booktabs}
\usepackage{multirow}

\shorttitle{Galaxy-Galaxy lensing in SDSS}
\shortauthors{Luo et al.}

\begin{document}

%%%%%%%%%%%%%%%%%%%%%%%%%%%%%%%%%%%%%%%%%%%%%%%%%%%%%%%%%%%

\title{Galaxy-Galaxy Weak-Lensing Measurements from SDSS: II. Host
  Halo properties of galaxy groups}

\author{Wentao Luo\altaffilmark{1}, Xiaohu Yang\altaffilmark{1,2},
  Tianhuan Lu\altaffilmark{3}, Feng Shi\altaffilmark{4}, Jun
  Zhang\altaffilmark{1}, H.J. Mo\altaffilmark{5,6},Chenggang Shu\altaffilmark{7}, Liping Fu\altaffilmark{7}, Mario Radovich\altaffilmark{8,9},Jiajun Zhang\altaffilmark{1}, Nan Li\altaffilmark{10}, Tomomi Sunayama\altaffilmark{11}, Lei Wang\altaffilmark{12}}

\altaffiltext{1}{Department of Astronomy, Shanghai Key Laboratory for
  Particle Physics and Cosmology, Shanghai Jiao Tong University,
  Shanghai 200240, China; Email: wentao.luo82@sjtu.edu.cn}

\altaffiltext{2}{IFSA Collaborative Innovation Center, and Tsung-Dao
  Lee Institute, Shanghai Jiao Tong University, Shanghai 200240,
  China; E-mail: xyang@sjtu.edu.cn}

\altaffiltext{3}{Zhiyuan College, Shanghai Jiao Tong University,
  Shanghai 200240, China}

\altaffiltext{4}{Shanghai Astronomical Observatory, Nandan Road 80,
  Shanghai 200030, China}

\altaffiltext{5} {Department of Astronomy, University of
  Massachusetts, Amherst MA 01003-9305, USA}

\altaffiltext{6} {Physics Department and Center for Astrophysics,
  Tsinghua University, Beijing 10084, China}
  
\altaffiltext{7} { Shanghai Key Lab for Astrophysics, Shanghai Normal
  University, 100 Guilin Road, 200234, Shanghai, China}

\altaffiltext{8} {INAF-Osservatorio Astronomico di Napoli, via
  Moiariello 16, I-80131 Napoli, Italy}

\altaffiltext{9} {INAF-Osservatorio Astronomico di Padova, vicolo
  dell'Osservatorio 5, I-35122 Padova, Italy}

\altaffiltext{10} {Centre for Astronomy \& Particle Theory at the 
  University of Nottingham, Nottingham NG7 2RD, UK}
 \altaffiltext{11}{Kavli Institute for the Physics and Mathematics of the Universe,
The University of Tokyo
5-1-5 Kashiwanoha, Kashiwa, Chiba 277-8583, Japa}
\altaffiltext{12}{Purple Mountain Observatory CAS, 8 Yuanhua Road, Nanjing,
  210023, China}

\begin{abstract}
  As the second paper of a series on studying galaxy-galaxy lensing
  signals using the Sloan Digital Sky Survey Data Release 7 (SDSS
  DR7), we present our measurement and modelling of the lensing
  signals around groups of galaxies. We divide the groups into four
  halo mass bins, and measure the signals around four different
  halo-center tracers: brightest central galaxy (BCG),
  luminosity-weighted center, number-weighted center and X-ray peak
  position.  For X-ray and SDSS DR7 cross identified groups, we
  further split the groups into low and high X-ray emission
  subsamples, both of which are assigned with two halo-center tracers,
  BCGs and X-ray peak positions. The galaxy-galaxy lensing signals
  show that BCGs, among the four candidates, are the best halo-center
  tracers. We model the lensing signals using a combination of four
  contributions: off-centered NFW host halo profile, sub-halo
  contribution, stellar contribution, and projected 2-halo term. We
  sample the posterior of 5 parameters i.e., halo mass, concentration,
  off-centering distance, sub halo mass, and fraction of subhalos via
  a MCMC package using the galaxy-galaxy lensing signals. After taking
  into account the sampling effects (e.g. Eddington bias), we found
  the best fit halo masses obtained from lensing signals are quite
  consistent with those obtained in the group catalog based on an
  abundance matching method, except in the lowest mass bin.
\end{abstract}

%%%%%%%%%%%%%%%%%%%%%%%%%%%%%%%%%%%%%%%%%%%%%%%%%%%%%%


\section{Introduction}
\label{sec_intro}

Modern galaxy formation theory suggests that dark matter halos form
first, which provide gravitational potential for galaxies to form. The
mass of dark matter halo is thus a critical quantity to understand
 galaxy formation and constrain the cosmological parameters.
There are many methods that can be used to constrain halo mass in
observations.  The luminosity of X-ray emission from the Intra Cluster
Medium (ICM) due to the thermal bremsstrahlung is scaled with the dark
matter halo mass assuming an hydro equilibrium state of the gas
\citep{Pratt2009}. 
%But this measurement is affected when there is
%merging process, or AGN feedback, both of which can cause turbulence
%and hence the equilibrium assumption no longer
%holds. 
The Sunyaev-Zeldovich \citep{SZ1972} effect, i.e.  the inverse
Compton scattering between the high energy electrons from clusters and
the CMB photons, is another indicator of cluster mass
\citep{Bleem2015}. 
%However, due to the projection effect and detection
%efficiency, this measurement is only applicable for a small number of
%massive clusters.

Optically, from large photometric galaxy surveys, clusters are
selected and assigned with dark matter halo masses in large sky
coverages and large redshift ranges.  Based on SDSS, SDSS-C4
\citep{Miller2005} and RedMaPPer \citep{Rykoff2014} select clusters
using photometric data alone, whereas MaxBCG sample
\citep{Koester2007} adds spectroscopic redshift as extra information
in selection criteria. All of those samples use an empirical scaling
relation between the effective number of member galaxies (richness)
and halo mass, a.k.a richness-mass relation, to assign halo masses to
clusters. Such methods are mainly applicable for those very massive
clusters and need to be scaled with additional measurements.

Based on spectroscopic redshift surveys, galaxy groups can be
extracted from halos of much lower mass either by the traditional
Friends-Of-Friends (FOF) method \citep[e.g.][]{Eke2004}, or by
sophisticated adaptive halo-based group finder\citep[e.g.][hereafter
Y07]{Yang2005, Yang2007}, which is more reliable in their membership
determination. From the group catalogs, a wide variety of mass
estimation methods are developed.  Y07 applied a luminosity ranking
and stellar mass ranking method to obtain halo mass estimation. For
poor systems, \cite{Lu2015} introduced an empirical mass estimation
method applying the gap between BCG luminosity and satellites
luminosity.  Assuming a Gaussian velocity distribution, satellite
kinematics \citep{Bosch2004} measure the dynamical mass of galaxy
groups. Similarly galaxy infall kinematics \citep[GIK]{Zu2013,Zu2014}
can also be used to constrain the halo mass.

Weak gravitational lensing signal, though statistical in nature, is
considered to be another powerful tool to study the property of dark
matter distributions due to the fact that the signal is sensitive to
all the intervening mass between the observer and the source.  The
successful measurement of weak lensing signals require high quality
imaging of background galaxies.  Many recent surveys like CFHTLens
\citep{Heymans2012}, DES \citep{Jarvis2015}, and KIDS
\citep{Kuijken2015, Viola2015}, etc., take weak lensing as one of
their key scientific projects. Future large surveys, either
ground-based or space-based such as EUCLID \citep{Refregier2010} and
LSST \citep{LSST2009} also take weak lensing as one of their key
projects. The high quality and deep galaxy images in terms of galaxy
number per square arc minutes enables one to measure the second order
weak lensing studies, a.k.a cosmic shear \citep{Fu2008,
  Kilbinger2013}, which are now constantly used to constrain the
cosmological parameters such as $\sigma_8$ and $\Omega_m$.

Sloan Digital Sky Survey \citep{York2000} initiates the {\it
  galaxy-galaxy} lensing analysis since \cite{Fischer2000}.  It is
followed by \cite{Sheldon2004}, who chose MaxBCG sample as lenses
binned by richness. \cite{Hirata2003} and \cite{Mandelbaum2005,
  Mandelbaum2006} not only studied the halo properties of lenses from
SDSS main sample, but also analysed sources of systematics caused by
PSF, selection effect, noise rectification effect and claimed that the
final signal should subtract the signal from random
sample. \cite{Simet2016} used redMaPPer as lenses and tested the
consistency of the mass-richness relation.  These galaxy-galaxy
lensing signals provided us another method to measure the halo mass of
lens systems in consideration.

To obtain reliable galaxy-galaxy lensing measurements, accurate image
processing of source galaxies are essential.  Many groups have
developed image processing pipelines devoted to improving the accuracy
of shape measurements \citep{Kaiser1995, Bertin1996, Maoli2000,
  Rhodes2000, vanWaerbeke2001, Bernstein2002,Bridle2002,
  Refregier2003, Bacon2003, Hirata2003, Heymans2005, Zhang2010,
  Zhang2011,Bernstein2014, Zhang2015}.  Among these, Lensfit
\citep{Miller2007, Miller2013, Kitching2008} applies a Bayesian based
model-fitting approach; BFD (Bayesian Fourier Domain) method
\citep{Bernstein2014} carries out Bayesian analysis in the Fourier
domain, using the distribution of un-lensed galaxy moments as a prior,
and the Fourier\_Quad method developed by \citep{Zhang2010, Zhang2011,
  Zhang2015, Zhang2016} uses image moments in the Fourier Domain.

Very recently, \cite{Luo2017} developed an image processing pipeline
and applied it to the SDSS DR7 imaging data. The galaxy-galaxy lensing
signals were measured for lens galaxies that are separated into
different luminosity bins and stellar mass bins.  As the second paper
of a series, here we measure the galaxy-galaxy lensing signals using
the group catalogs constructed by \cite{Yang2007}. We will measure the
galaxy-galaxy lensing signals around groups in different mass bins, as
well as for four types of halo-center tracers: brightest central
galaxy (BCG), luminosity-weighted center (LwCen), number-weighted
center (NwCen) and X-ray peak position. We focus on the estimation of
halo properties of the groups in consideration such as their halo
masses, concentrations and off-center effects, etc.

The structure of this paper is organized as follows.  In section
\ref{data}, we first present the data set and the main results of
galaxy-galaxy lensing signals measured around galaxy groups.  We show
our modelling and parameter constraints in section \ref{model}. We
discuss the Eddington bias that the group mass estimation may have in
section \ref{sec_bias}. We summarize our results in section
\ref{summary}. Unless stated otherwise, we adopt a LCDM cosmology with
$\Omega_m=0.28$, $\Omega_{\Lambda}=0.72$ and $h=1.0$.

\begin{figure*}
\centering
\includegraphics[width=7.5cm,height=7.5cm]{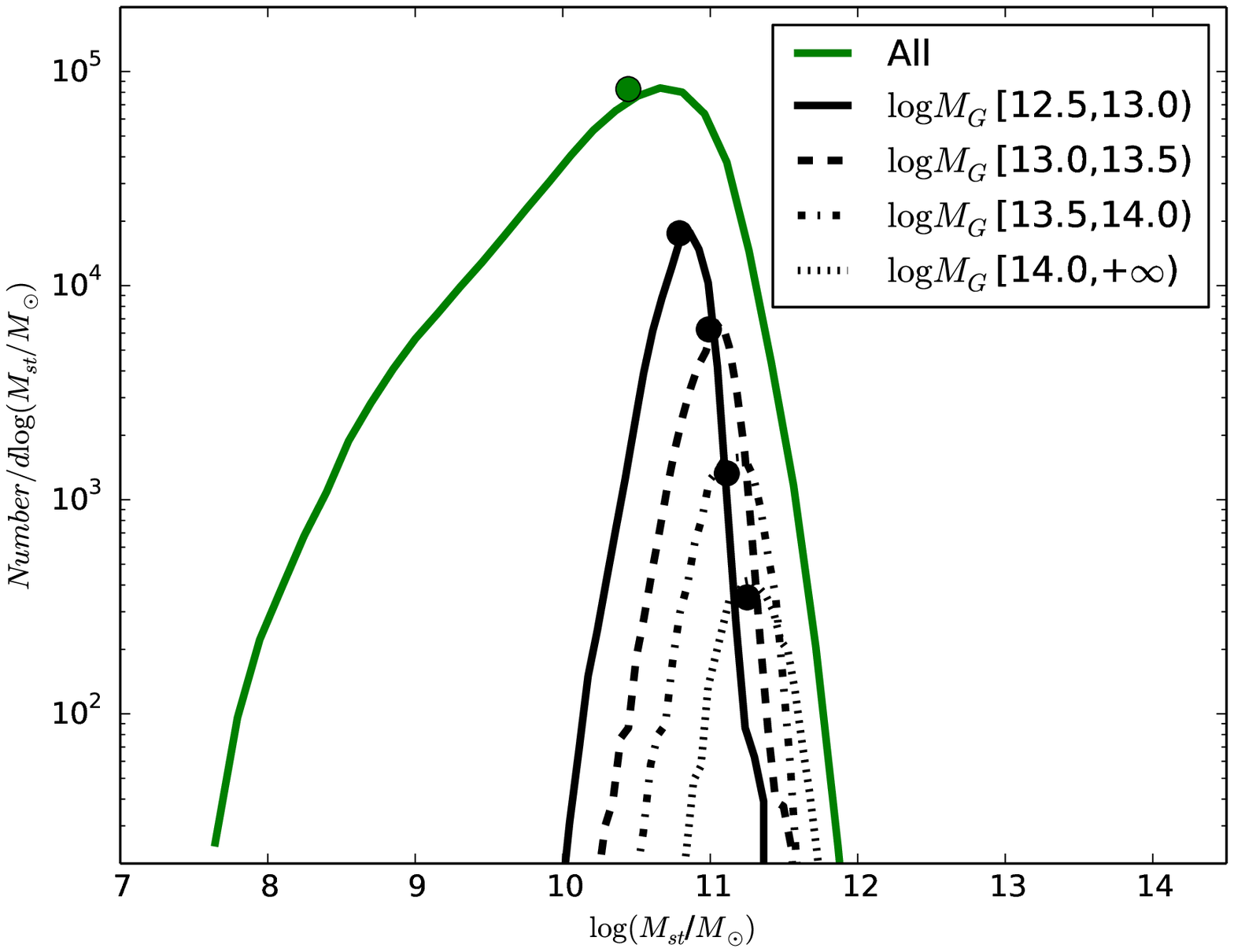}
\includegraphics[width=7.5cm,height=7.5cm]{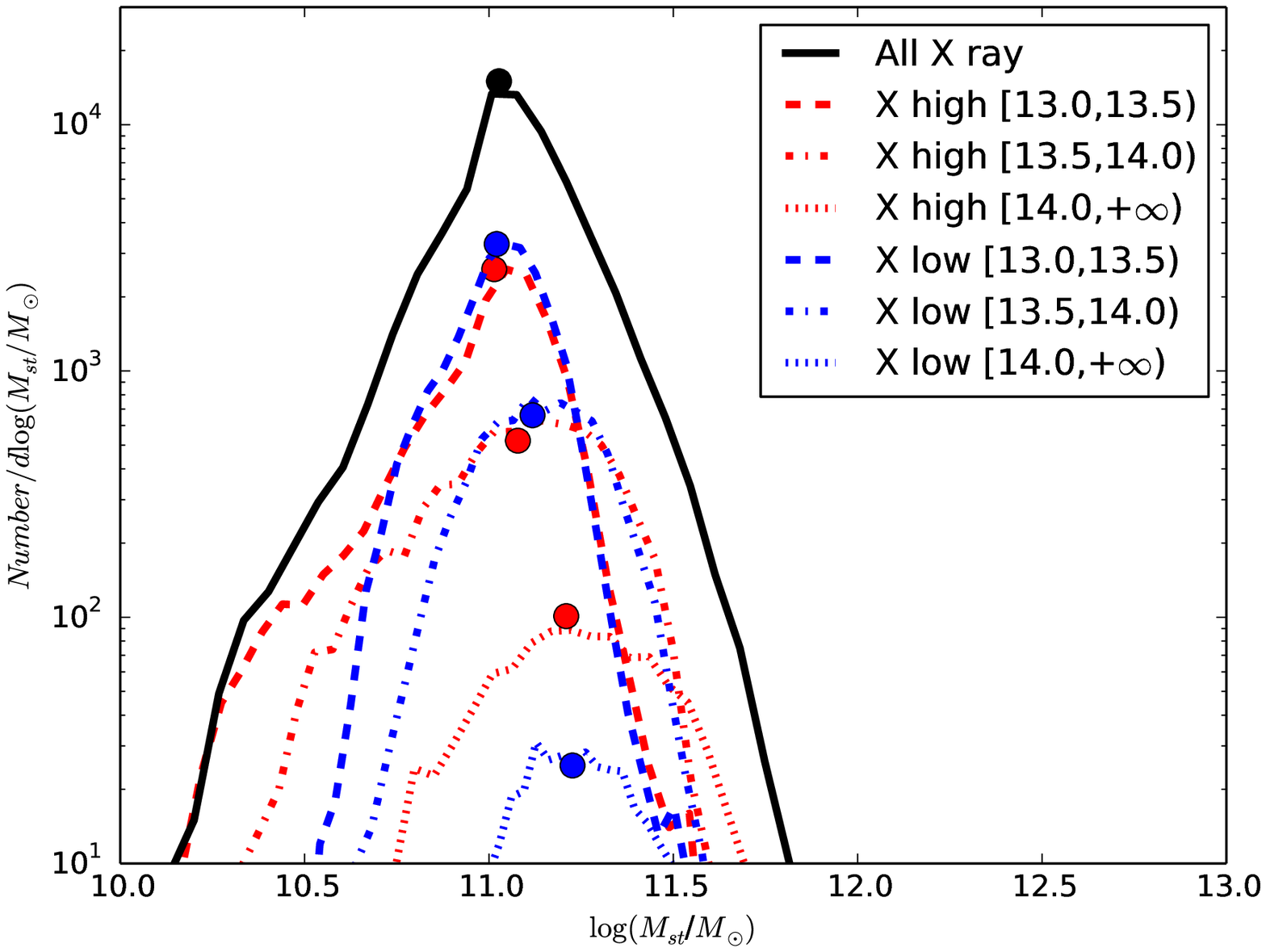}
\caption{Left panel: The green solid line and the green dot denote the
  stellar mass distribution and the average stellar mass of all the
  galaxies. The black solid, dashed, dot-dashed, dotted lines and the
  black dots are the stellar mass distributions and average stellar
  masses of BCGs in different halo mass bins as indicated.  Right
  panel: Similar to the left panel, but here for high (red) and low
  (blue) X-ray luminosity subsamples. }
  \label{fig:stellar}
\end{figure*}

\section{The Galaxy-Galaxy Lensing Signals of Galaxy Groups}
\label{data}

\subsection{Sources}
\label{source}

We use the shape catalog created by \cite{Luo2017} based on SDSS DR7
imaging data. The DR7 imaging data, with \textit{u}, \textit{g},
\textit{r}, \textit{i} and \textit{z} band, covers about 8423 square
degrees of the LEGACY sky ($\sim$230 million distinct photometric
objects). The total number of objects identified as galaxies is around
150 million.  These galaxies are further selected and processed by our
image pipeline. The galaxies are selected by OBJC\_TYPE=3 from PHOTO
pipe \citep{Lupton2001} and must be brighter than 22 in $r$ band model
magnitude and brighter than 21.6 in $i$ band model magnitude.  We also
apply Flags such as BINNED1 (detected at $\geq 5$), SATURATED=0 (do
not have saturated pixels), EDGE=0 (do not locate at the edge of the
CCD), MAYBE\_CR=0 (not cosmic rays), MAYBE\_EGHOST=0 (not electronic
ghost line) and PEAKCENTER=0 (centroiding algorithm works well for
this subject). In processing the images, the Point Spread Function
(PSF) effect was corrected by combining \cite{Bernstein2002} and
\cite{Hirata2003} method.  After the shape measurement, a size cut by
resolution factor (${\cal R}\geq 1/3$) was applied. The final shape
catalog for our study contains 41,631,361 galaxies with position,
shape, shape error and photoZ information.

\subsection{Lenses}
\label{lens}

The galaxies used in our probe are obtained from the New York
University Value-Added Galaxy Catalogue
\citep[NYU-VAGC;][]{Blanton2005}, which is based on SDSS DR7
\citep{Abazajian2009} but with an independent set of significantly
improved reductions.  From the NYU-VAGC, we select all galaxies in the
Main Galaxy Sample with an extinction corrected apparent magnitude
brighter than $r=17.72$, with redshifts in the range
$0.01 \leq z \leq 0.20$ and with a redshift completeness
${\cal C}_z > 0.7$.  This gives a sample of $639,359$ galaxies with a
sky coverage of 7748 square degrees.  For each galaxy, we estimate its
stellar mass using the fitting formula of \cite{Bell2003}.

From this galaxy catalog, a total of 472,113 groups are selected using
the halo based group finder \citep{Yang2007}, each of which has been
assigned with a halo mass by the ranking method.  With the halo mass
information, we bin the groups with mass $\ge 10^{12}M_{\odot}$ into
four mass bins. Shown in the left panel of Fig.\ref{fig:stellar} are
the stellar mass distribution of all the galaxies (green solid line)
and BCGs in different halo mass bins as indicated. The related average
values of these galaxies are shown as the solid dots on top of the
lines. There is a monotonic increase of average stellar mass of BCGs
for the 4 mass bin groups. Tab. \ref{tab:tbl-1} lists some
properties of the groups in these four mass bins. For each mass bin,
we define four types of halo center tracers i.e. BCG, luminosity
weighted center (LwCen), number weighted center (NwCen) and X-ray peak
position (no X-ray measurement for the first mass bin).

\begin{table}
\begin{center}
  \caption{\label{tab:tbl-1} Properties of the four lens samples
    created for this paper.} 
\begin{tabular}{cccccc}
  \hline\hline \\
  Sample & $\log M_G$ & $N_{grp}$ & $\langle z \rangle $ & $\log \langle 
         M_{st} \rangle$ & $\log \langle M_G\rangle$ \\  \\
  \hline \\
  M1  & $(12.5,13.0]$  & 101042 & 0.13 & 10.77 & 12.72 \\
  M2  & $(13.0,13.5]$  & 43896  & 0.15 & 10.97 & 13.21 \\
  M3  & $(13.5,14.0]$  & 14707  & 0.15 & 11.10 & 13.70 \\
  M4  & $(14.0,\infty]$& 4033   & 0.15 & 11.25 & 14.25 \\ \\
  \hline
\end{tabular}

\tablecomments{Columns 1-6 correspond to the sample name, group mass
  range, number of groups, average redshift, stellar mass and group
  mass, respectively.}

\end{center}
\end{table}

\begin{table}
\begin{center}
  \caption{\label{tab:tbl-2} High/low X ray dichotomy. }
\begin{tabular}{cccccc}
\hline \hline \\
Mass bin & $L_X$ & $N_{grp}$ & $\langle z \rangle $ & $\log \langle 
         M_{st} \rangle$ & $\log \langle M_G \rangle$ \\ \\
\hline \\
\multirow{2}{*}{13.0-13.5} & high & 14582 & 0.14 & 10.97 & 13.24 \\
  & low & 22086 & 0.15 & 11.00 & 13.23  \\ \\
\hline \\
\multirow{2}{*}{13.5-14.0} & high & 7285 & 0.14 & 11.05 & 13.71\\
  & low & 7049 & 0.16 & 11.12 & 13.68\\ \\
\hline \\
\multirow{2}{*}{14.0-above} & high & 2953 & 0.15 & 11.20 & 14.27\\
  & low & 1080 & 0.16 & 11.23 & 14.18 \\ \\
\hline
\end{tabular}

\tablecomments{Columns 1-6 correspond to group mass range, X-ray
  luminosity status, number of groups, average redshift, stellar mass
  and group mass, respectively.}

\end{center}
\end{table}

For the three of our halo bins with halo mass $\log M_G\ge 13.0$,
\cite{Wang2014} measured their X-ray luminosities using the ROSAT
data.  We divide groups in each of the mass bin into high X-ray
luminosity and low X-ray luminosity subsamples with roughly equal
numbers using the $L_X$ parameter provided in \cite{Wang2014} catalog,
where $L_X$ is the X-ray luminosity of a group in 0.2-2.4kev in unit
of $10^{44}$ erg/s. The related information of these subsamples are
listed in Tab.  \ref{tab:tbl-2}.  There is a slight difference between
the mean redshift of the low X-ray luminosity and high X-ray
luminosity subsamples. Both high and low X-ray luminosity samples are
assigned with BCG and X-ray peak position as different halo center
tracers.  Note that since we are using the group masses estimated
using the ranking of characteristic group luminosity, while the ones
provide in \citet{Wang2014} are based on the ranking of characteristic
group stellar mass, some of our groups are not assigned with X-ray
luminosities. Thus the total number of groups used in the X-ray high
and low subsamples is slightly reduced, especially in the low mass
bin.

Shown in the right panel of Fig. \ref{fig:stellar} are the stellar
mass distributions of central galaxies associated with the X-ray peak
positions in different halo mass bins as indicated. The dots on top of
the lines are their average values. Here results are shown for low
X-ray luminosity (blue) and high X-ray luminosity subsamples (red),
respectively.  Again, we see a monotonic increase of average stellar
mass of central galaxies for the three mass bin groups.

\subsection{Measure the galaxy-galaxy lensing signals}

\begin{figure*}
\centering
\includegraphics[width=14cm,height=11cm]{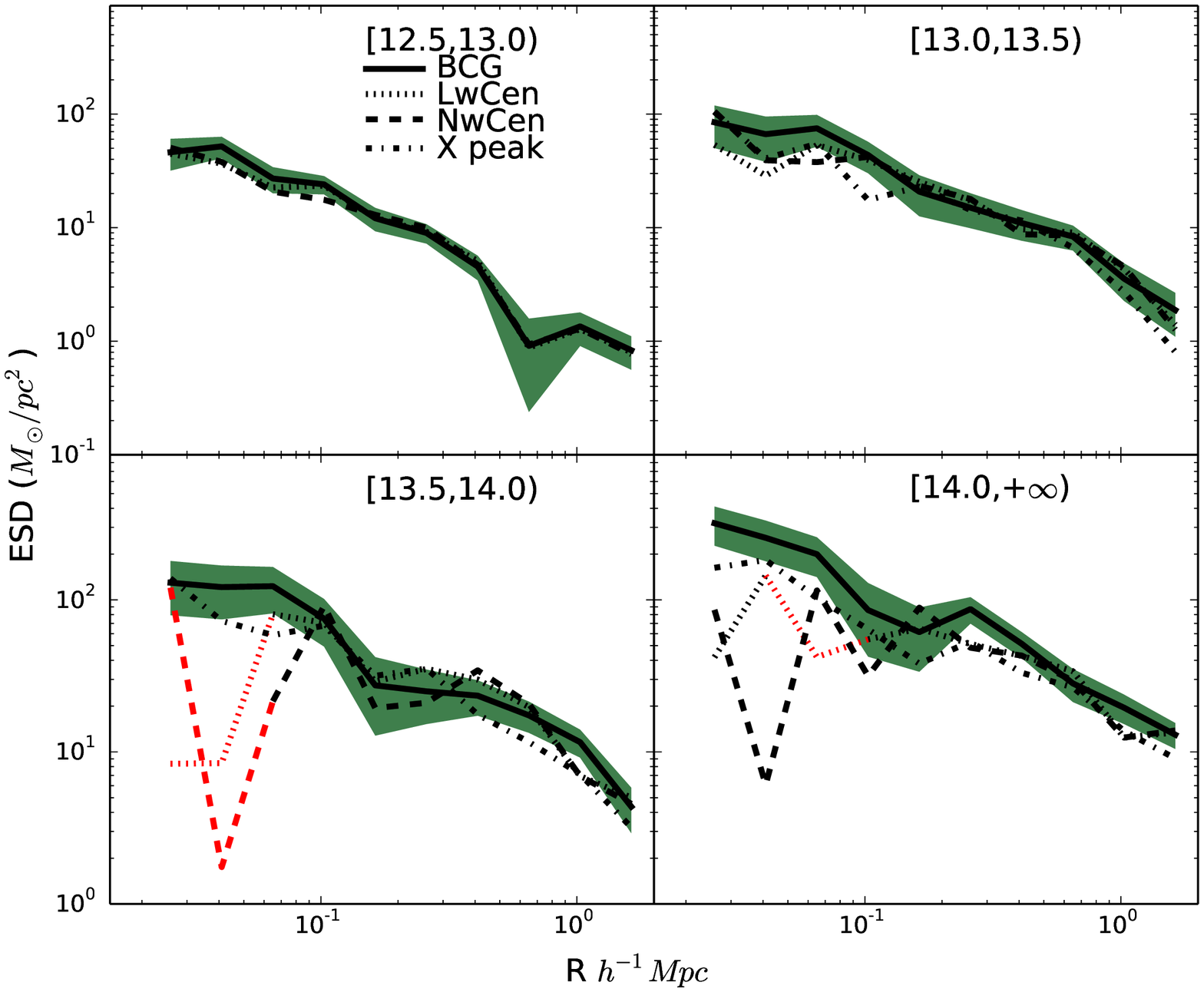}
\caption{The ESD profiles measured from 4 mass bins along with
  different halo center tracers. Solid lines with green shaded areas
  represent results for BCGs.  The dotted, dashed, dot-dashed lines
  are results for LwCens, NwCens and X-ray peak positions,
  respectively. Red parts denote negative values.}
  \label{fig:massbins}
\end{figure*}

\begin{figure*}
\centering
\includegraphics[width=14cm,height=14cm]{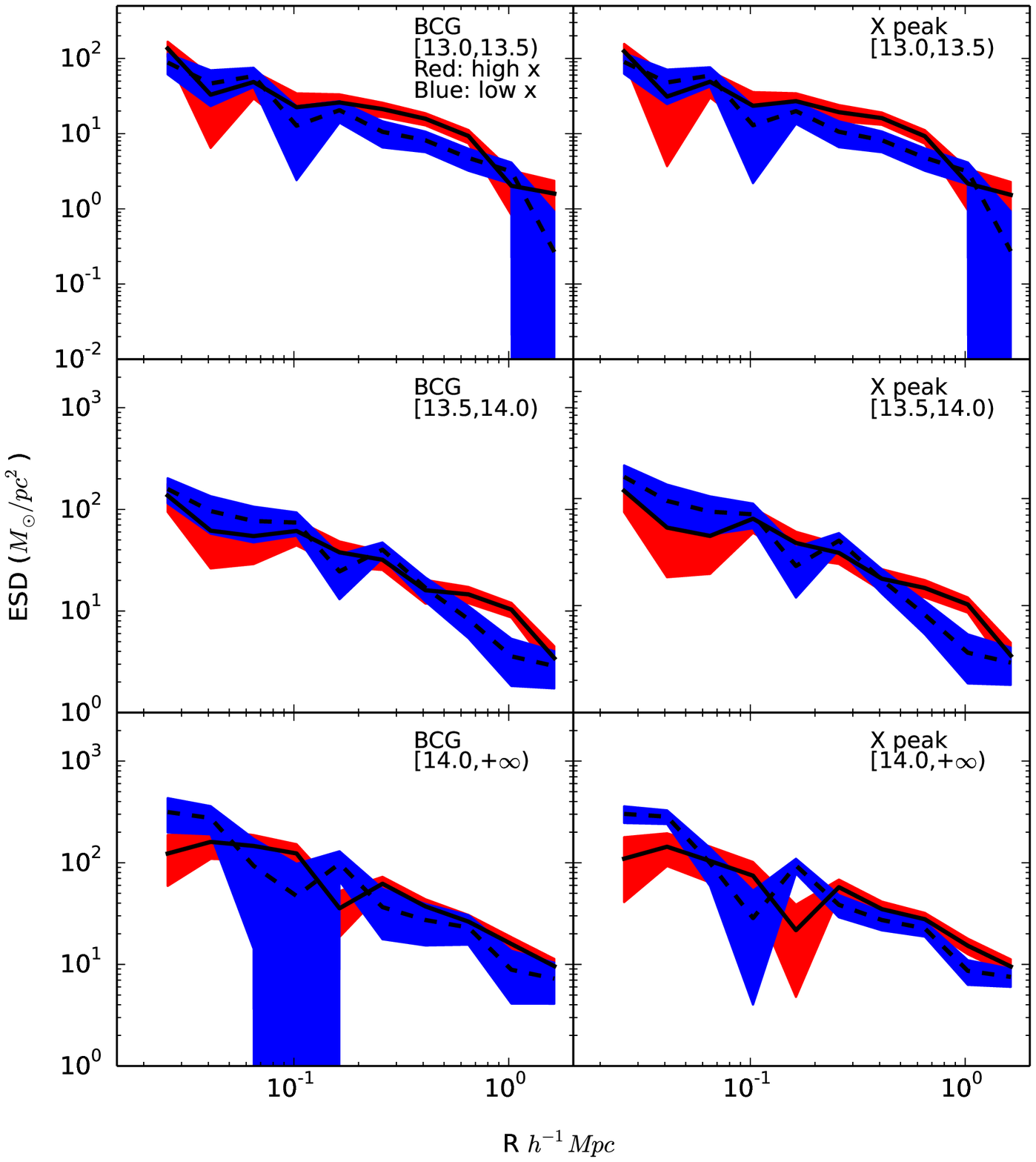}
\caption{The ESD profiles measured from X-ray high and low luminosity
  subsamples. From top panel to bottom panel, the mass increases as
  indicated in each panel. Black solid lines with red one sigma
  regions are the ESD profiles measured from X-ray luminous subsamples
  while dashed lines with blue one sigma regions denote the X-ray
  faint subsamples.  Left and right columns are results measured
  around BCGs and X-ray peak positions, respectively.}
  \label{fig:xrays}
\end{figure*}

The shear signals $\gamma$ along any desired directions can be
measured by the weighted mean of source galaxy shapes,
\begin{equation}
\gamma_l=\frac{1}{2\bar{R}}\frac{\sum w_ie_l^{\rm rot}}{\sum w_i}\,,
\end{equation}
where $l=1,2$ and $w_i$ is a weighting function. Here $\bar{R}$ is the
mean responsivity of our survey galaxies which is defined as
\begin{equation}
\bar{R}\equiv 1- {1\over N} \sum_{i=1}^{N} (e_1^{\rm rot})^2 \,,
\end{equation}
where $N$ is the total number of source images and $e_1^{rot}$ is the
ellipticity of background galaxies after SPA (the angle between the
camera column position with respect to north from fpC files) rotation.
The weighting term is composed of two components,
\begin{equation}
w=\frac{1}{\sigma_{\rm sky}^2+\sigma_{\rm shape}^2}\,,
\end{equation}
where $\sigma_{\rm shape}=<e^2>$ is the shape noise and
$\sigma_{\rm sky}=\frac{\sigma^{pix}}{{\cal R}F}\sqrt{4\pi n}$ is the
sky noise. Here $\sigma^{pix}$ denotes the size of galaxy in pixels,
${\cal R}$ is the resolution factor, $F$ is the flux and $n$ the sky
and dark current in ADU.

The tangential shear of a lens system is connected to its Excess
Surface Density (ESD) by the geometry factor
$\Sigma_{crit}(z_l,z_s)=\frac{c^2}{4\pi G}\frac{D_s}{D_lD_{ls}}$
\begin{equation}
\label{eq:esd}
\Delta\Sigma(R) =\Sigma(\leq
R)-\Sigma(R)=\gamma_t\Sigma_{crit}(z_l,z_s) \,.
\end{equation}
Here, $z_l$ and $z_s$ denotes the redshifts of the group and the
source respectively. $D_l$, $D_s$ and $D_{ls}$ are the angular
diameter distance of the lens, the source and between the lens and the
source. $\Sigma(\leq R)$ is the average surface density inside the
projected distance $R$, and $\Sigma(R)$ is the surface density at the
projected distance $R$.

Shown in Fig. \ref{fig:massbins} are the ESD profiles measured around
groups in different halo mass bins as indicated in each panel. In each
panel, results are shown for different halo center tracers: BCG (black
solid lines with one sigma error as green band), LwCen (black dotted
line), NwCen (black dashed line) and X-ray peak position (black dash
dotted line).  Here the error bars are obtained by 2000 bootstrap
resampling of the lens systems in consideration.  The negative values
from the measured ESDs are denoted by red color.  At smaller scale
$<100\kpch$, for more massive bins, the ESD signals around LwCen and
NwCen begin to deviate from that of BCGs. Among these four types of
halo center tracers, BCGs have the steepest ESD profiles at small
scales, suggesting that BCGs are the best halo-center tracers.

Shown in Fig. \ref{fig:xrays} are the ESDs measured from X-ray high
and low luminosity groups.  The solid curves with red one sigma error
bands are the signals measured around high X-ray luminosity subsamples
while the dashed lines with blue one sigma bands are results for low
X-ray luminosity subsamples. There are some differences between the
ESDs of high and low X-ray luminosity groups. In massive groups with
mass $\ga 10^{13.5}M_{\odot}$, the low X-ray luminosity groups show
somewhat more prominent ESDs at small scales, while in less massive
groups the dependance is opposite. At large scales, the high X-ray
luminosity groups show overall higher amplitude of ESDs indicating
that the masses of their host halos are somewhat more massive.

\section{The halo properties of galaxy groups}
\label{model}

With the ESDs measured for different group samples and subsamples in
previous section, we proceed to constrain the related halo properties
of galaxy groups in this section.

\subsection{Weak Lensing Model}

\begin{figure*}
\centering
\includegraphics[width=7.5cm,height=7.5cm]{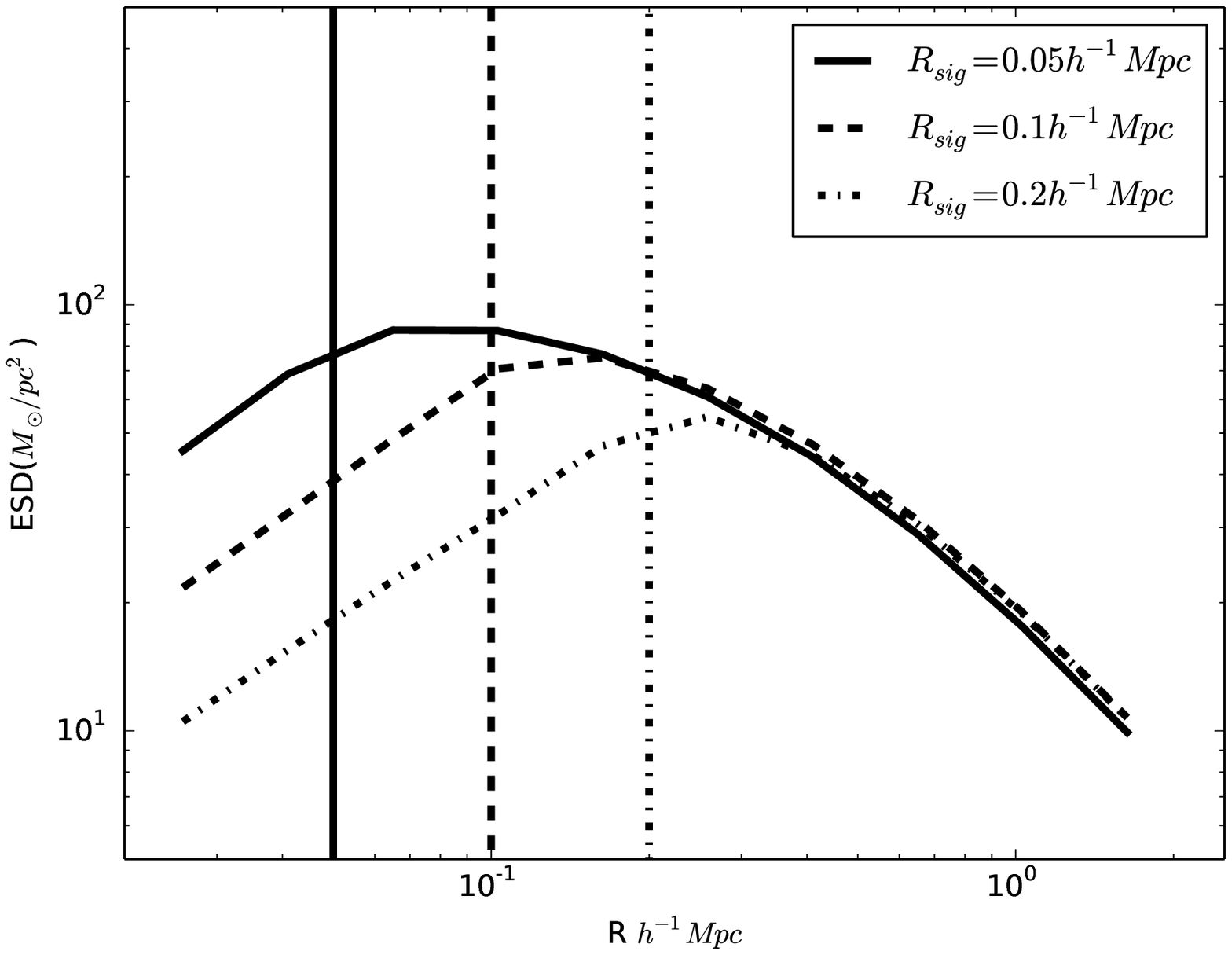}
\includegraphics[width=7.5cm,height=7.5cm]{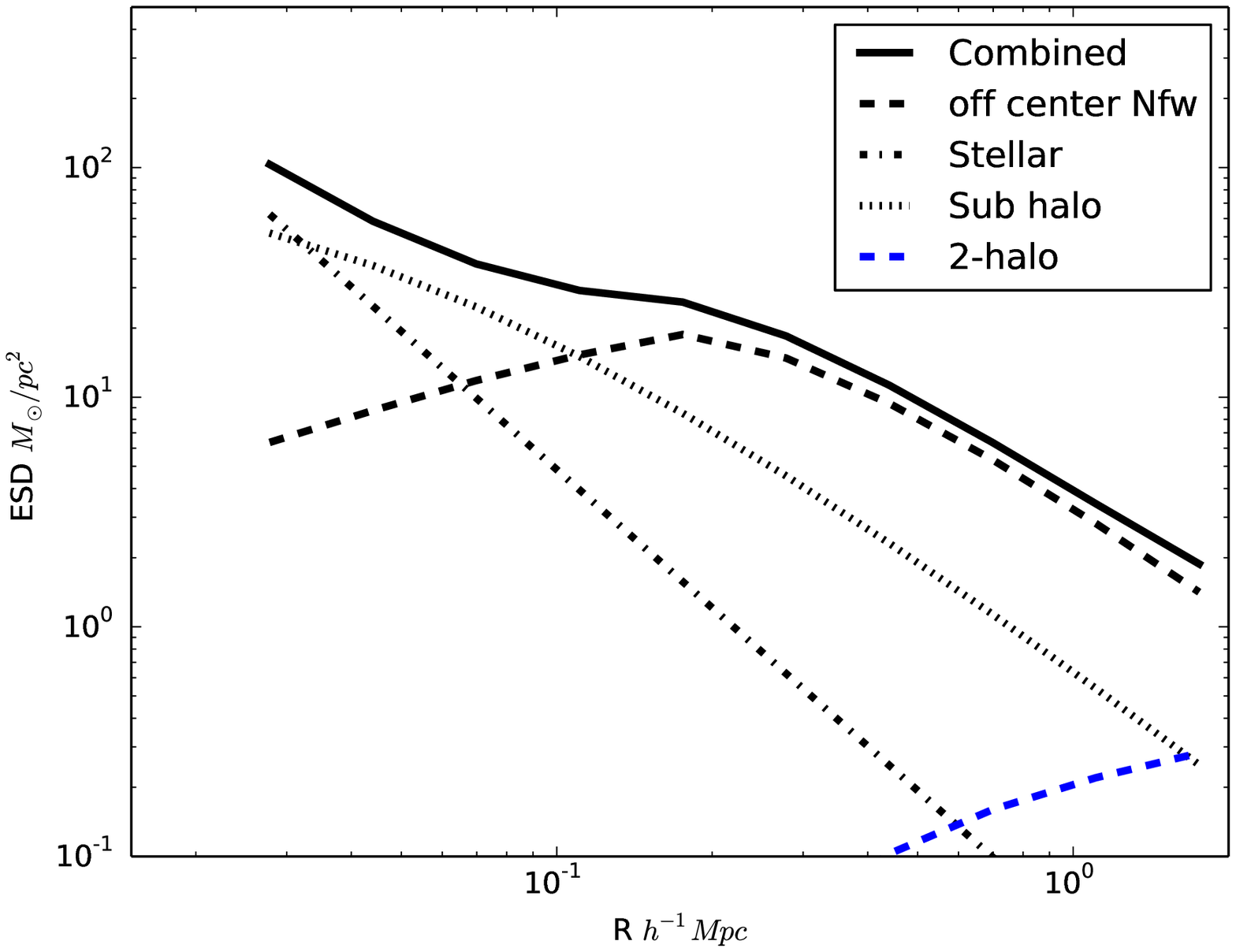}
\caption{Left panel: Black line is the theoretical prediction of ESD
  profile with $R_{\rm sig} =0.05\mpch$, while the dashed line and the
  dot-dashed line are the ESD profiles with $R_{\rm sig}$=0.1 and
  $0.2\mpch$, respectively.  Two vertical lines denotes the
  $R_{\rm sig}$ values. Right panel: The model we adopt to describe
  the lensing ESD profile. The dashed, dotted, dot-dashed and blue
  dashed lines are the contributions from an off-centered NFW with
  $R_{\rm sig}=0.1\mpch$, the subhalo, the stellar mass of galaxy in
  consideration and the 2-halo term, respectively.  The black solid
  line is the combined profile.}
  \label{fig:esdmodel}
\end{figure*}

The ESD around a lens galaxy is related to the line-of-sight
projection of the galaxy-matter cross correlation function,
\begin{equation}
\xi_{\rm gm}(r)=\langle\delta({\bf x})_{g}\delta({\bf x}+{\bf r})_{m}\rangle, 
\end{equation}
so that
\begin{equation}
\label{sigatr}
\Sigma(R) = 2 \overline{\rho} \int_{R}^{\infty} \xi_{\rm gm}(r) 
{r \, \rmd r \over \sqrt{r^2 - R^2}}\,,
\end{equation}
and
\begin{equation}
\label{siginr}
\Sigma(\leq R) = \frac{4\overline{\rho}}{R^2} \int_0^R y\,\,dy\,
 \int_{y}^{\infty} \xi_{\rm gm}(r) {r \, \rmd r \over \sqrt{r^2 - y^2}}\,,
\end{equation}
where $\overline{\rho}$ is the average background density of the
Universe.  Note that in both equations, we have omitted the
contribution from the mean density of the universe, as it does not
contribute to the ESD.  In general, the ESD is composed of the
following four components: host halo mass, subhalo mass if it is a
satellite or interloper, the stellar mass associated with the galaxy
in consideration and projected two halo term,
\begin{equation}
  \Delta\Sigma(R)=\Delta\Sigma_{host}(R)+\Delta\Sigma_{sub}+
\Delta\Sigma_{*}(R)+\Delta\Sigma_{2h}\,.
\end{equation}

According to \cite{Yang2006a}, if the candidate lens galaxy (system)
locates at the center of host halo, the average projected density of
the host halo can be calculated from the NFW profile.  Assuming an NFW
profile of the host halo, we have
\begin{equation}
\rho(r)=\frac{\rho_0}{(r/r_s)(1+r/r_s)^2},
\end{equation} 
with $\rho_0=\frac{{\bar\rho\Delta_{vir}}}{3I}$, where
$\Delta_{vir}=200$, $I=\frac{1}{c^3}\int_0^c
\frac{xdx}{(1+x)^2}$. Here $c$ is the concentration parameter defined
as the ratio between the virial radius of a halo and its
characteristic scale radius $r_s$.  The projected surface density then
can be analytically expressed as \citep{Yang2006a}:
\begin{equation}
\label{eq:nfw}
\Sigma_{\rm NFW}(R)=\frac{M_h}{2\pi r_s^2I} f(x)\,,
\end{equation}
where $M_h$ is the halo mass and $f(x)$ bears the following form with
$x=R/r_s$:
\begin{equation}  
f(x)=
\left\{  
  \begin{array}{lr}  
   \frac{1}{x^2-2}[1-\frac{\ln{\frac{1+\sqrt{1-x^2}}{x}}}{\sqrt{1-x^2}}] & x<1  \\  
   \frac{1}{3} & x=1\\  
   \frac{1}{x^2-1}[1-\frac{atan(\sqrt{x^2-1})}{\sqrt{x^2-1}}] & x>1 \,.
  \end{array} 
\right. 
\end{equation}

\begin{figure*}
\centering
\includegraphics[width=14cm,height=14cm]{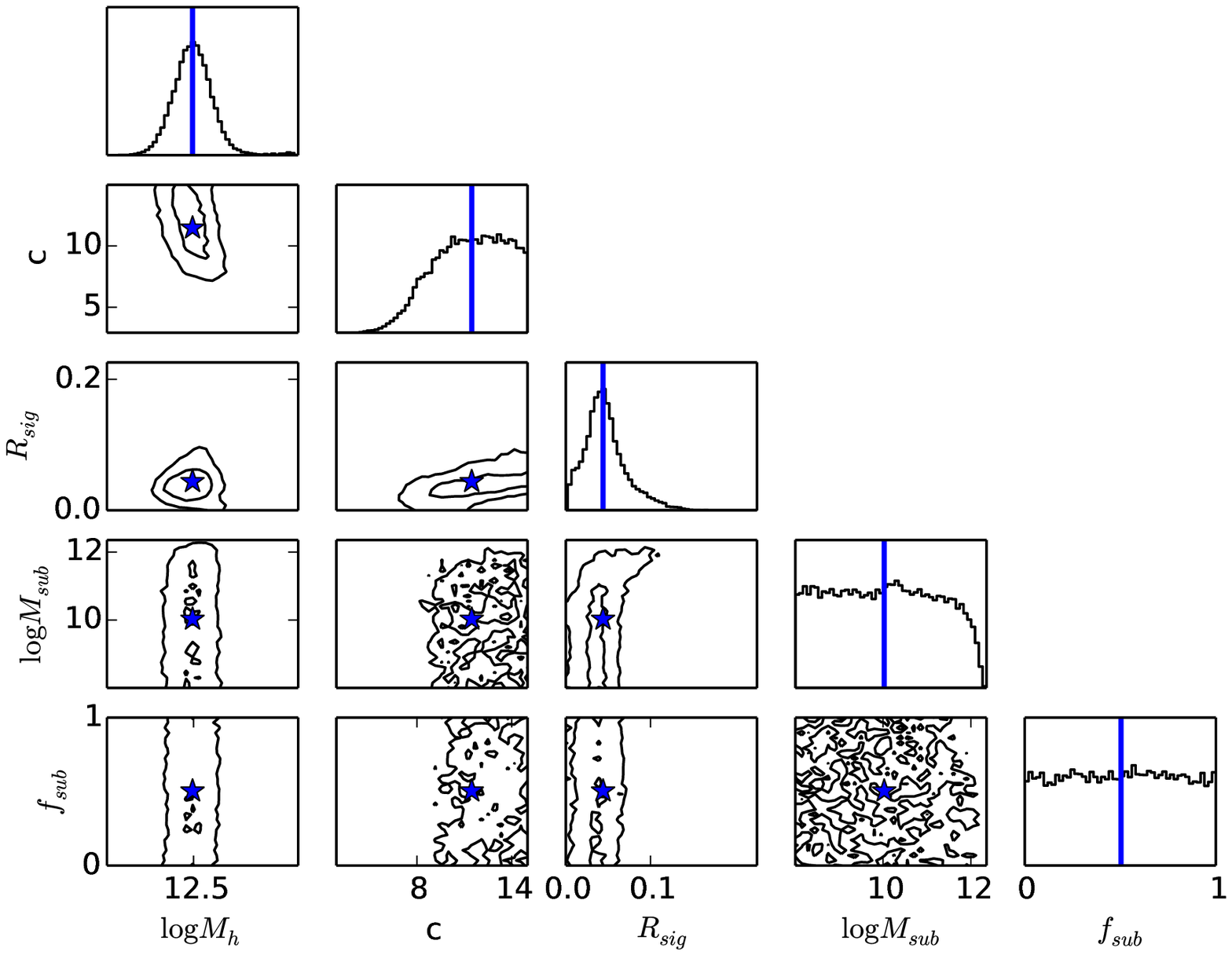}
\caption{The marginalized posterior distributions of the five free
  parameters for the lowest mass bin groups. The blue stars (blue
  solid lines) denote the median value of each distribution. The two
  contour levels correspond to the 68\% and 95\% confidence levels,
  respectively. }
  \label{fig:posterior}
\end{figure*}

On the other hand, if the candidate lens galaxy does not locate at the
center of the host halo, but with an off-center distance $R_{\rm off}$, the
projected surface density will change from an NFW profile
$\Sigma_{\rm NFW}(R)$ to
\begin{eqnarray}
&&\Sigma_{host}(R|R_{\rm
  off}) = \nonumber \\
&&\frac{1}{2\pi}\int_{0}^{2\pi}\Sigma_{\rm NFW}(\sqrt{R^2+R_{\rm
    off}^2+2R_{\rm off}Rcos\theta}) \, d\theta \,.
\end{eqnarray}
Here, we adopt the offset model proposed by \citet{Johnston2007},
where $R_{\rm off}$ follows a 2D Gaussian distribution.  This model
is drawn from the mock catalog based on ADDGALS technique
\citep{Wechsler2006} combined with light-cone from Hubble Volume
simulation \citep{Evrard2002}.  The resulting projected density
profile is the convolution between the $R_{\rm off}$ and the
$\Sigma_{host}(R|R_{\rm off})$,
\begin{equation}
\Sigma_{host}(R)=\int dR_{\rm off}P(R_{\rm off})\Sigma(R|R_{\rm off})\,,
\end{equation}
where
\begin{equation}
P(R_{\rm off})=\frac{R_{\rm off}}{R_{\rm sig}^2}\exp(-0.5(R_{\rm
  off}/R_{\rm sig})^2).
\end{equation}
Here $R_{\rm sig}$ is the dispersion of $P(R_{\rm off})$. Thus in
total, we have three free parameters regarding the host halo
properties, $M_h$, $c$ and $R_{\rm sig}$, to be constrained using the
observed ESDs.  As an illustration, we show in the left panel of
Fig. \ref{fig:esdmodel} how the measured ESDs may vary as a function
of $R_{\rm sig}$. Here we adopt a fixed halo mass and concentration
with $\log M_h=14.0$ and $c=7.0$ based on \cite{Zhao2009} formula, and
vary $R_{\rm sig}=0.05, 0.1, 0.2\mpch$, respectively.  Larger
$R_{\rm sig}$ moves the ESD peak further away from measurement center,
and suppress the signal at small scale.

Next, we consider the subhalo contribution. There is some possibility
that the candidate lens galaxy is not the true central galaxy and may
contain the subhalo component. In addition, in the group finder, there
are some possibilities that the central galaxy is an interloper and
may contain its original host halo component.  For these reasons, we
introduce a subhalo contribution in our ESD modelling. We assume that
a fraction of $f_{sub}$ contain subhalo with mass $\log M_{sub}$,
\begin{equation}
\Sigma_{sub}(R)=f_{sub}\Sigma_{\rm NFW}(R|M_{sub},c=15)\,.
\end{equation}
Here we simply fix the concentration as $c=15$\footnote{As subhalos
  are on average relatively low mass ones and may be affected by
  stripping effect, their concentration is thus set to relatively
  large values. Change this value to 10 does not impact any of our
  results significantly. } and treat $\log M_{sub}$ and $f_{sub}$ as
our fourth and fifth free parameters in our modelling. In addition, we
require that $\log M_h - \log M_{sub}\ge 0.3$ and $f_{sub}\le 1.0$.

Then we consider the stellar mass component. As pointed out in
\cite{Johnston2007} and \cite{George2012}, the stellar mass component
can be treated as a point mass, and the related ESD can be modelled
simply as,
\begin{equation}
\Delta\Sigma_{*}(R)=\frac{M_*}{2\pi R^2}\,,
\end{equation}
where $M_*$ is the stellar mass of candidate central galaxy in
consideration. We directly use the average stellar mass of galaxies as
given in Table \ref{tab:tbl-1} and \ref{tab:tbl-2} in our modelling.

Finally, for the signal caused by the 2-halo term, we calculate the
power spectrum at the mean redshift of each sample using the CAMB
(Code for Anisotropies from Microwave Background ) of \cite{Lewis2013}
and then convert the power spectrum to matter-matter correlation
function $\xi_{mm}$ \citep{Takahashi2012}. $\xi_{mm}$ is then related
to halo-matter correlation function $\xi_{hm}$ via halo bias
$b_h(M_h)$ model \citep{Seljak2004}
\begin{equation}
\xi_{hm}=b_h(M_h)\xi_{mm}.
\end{equation} 
To be more precise, we use the scale dependent bias model of \cite{Tinker2005},
\begin{equation}
\xi_{hm}=b_h\eta\xi_{mm}\,,
\end{equation}
where
\begin{equation}
\eta(r)=\frac{(1+1.17\xi_{mm}(r))^{1.49}}{(1+0.69\xi_{mm}(r))^{2.09}}.
\end{equation}
The 2-halo term projected mass density $\Delta\Sigma_{2halo}(R)$ is
then calculated using Eqs. \ref{eq:esd}, \ref{sigatr} and \ref{siginr}.
In practice, we only make the integration to a distance $50\mpch$ to
model the 2-halo term contribution \citep[see also][]{Niemiec2017}.

As an illustration, we show in the right panel of
Fig. \ref{fig:esdmodel} each components of the model. Here we adopt
$\log M_h=12.5, c= 10.0, \log M_{sub}=11.0$ and $f_{sub}=0.5$.  The
off-centered NFW is shown as the dashed line, stellar component as the
dot-dashed line, sub halo as dotted line and the 2-halo term as the
blue dashed line.  The combined signal is the black solid curve. From
this plot, we can see that at scales smaller than $50\kpch$, stellar
contribution becomes significant. The 2-halo term contribution is only
important at scales larger than a few virial radii.

\begin{figure*}
\centering
\includegraphics[width=14cm,height=11cm]{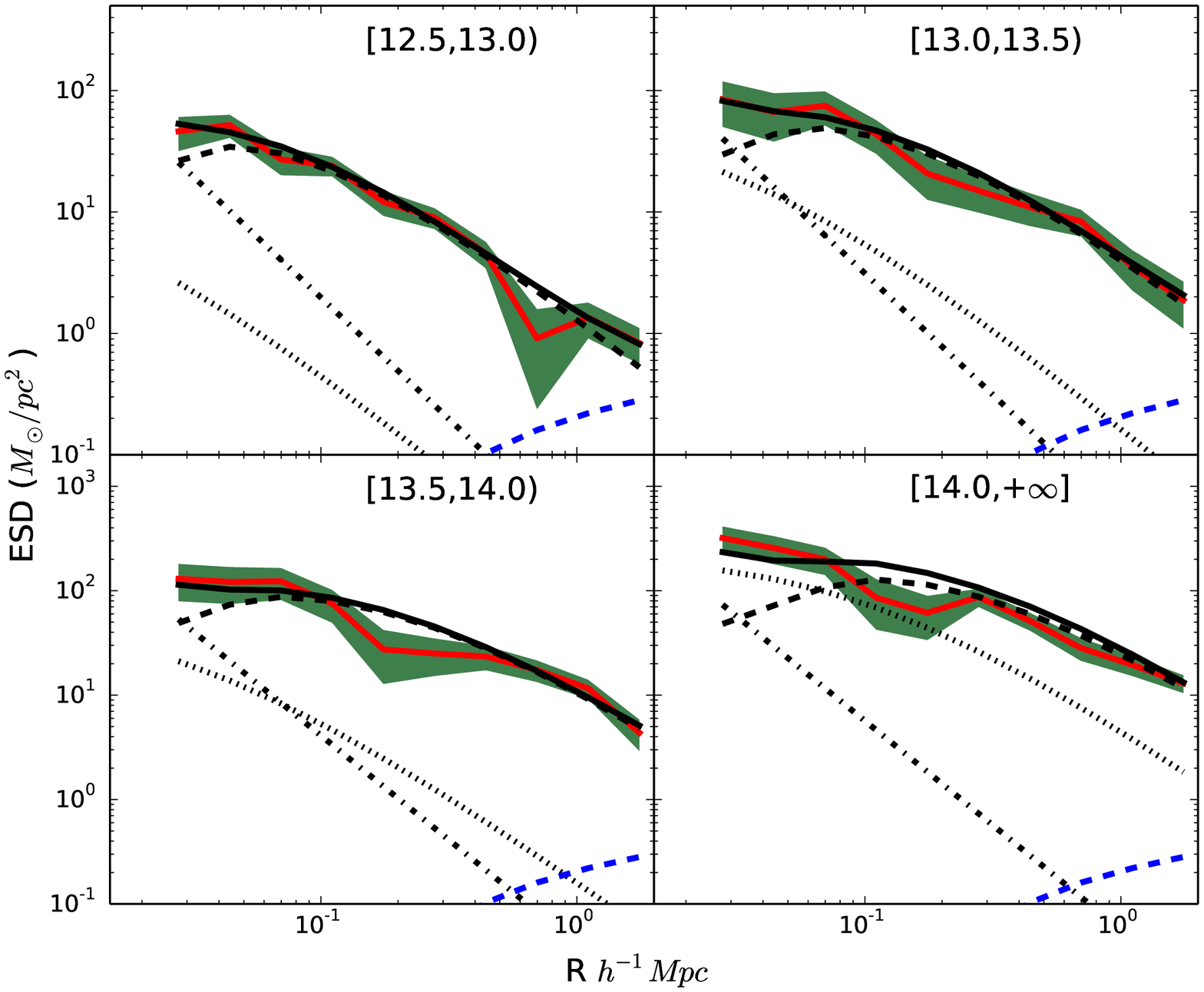}
\caption{The ESDs measurements (shaded area) and the best fit model
  predictions (black solid line) for groups in four different mass
  bins. The dashed, dotted, dot-dashed and blue dashed lines are the
  contributions from an off-centered NFW, the subhalo, the stellar
  mass of galaxy in consideration and the 2-halo term, respectively. }
  \label{fig:fitted}
\end{figure*}

\begin{table*}
\begin{center}
  \caption{\label{tab:tbl-3} Posterior of parameters we fitted to our
    group samples. }
\begin{tabular}{cccccccc }
  \hline \hline\\
  Sample & Mass bin & Centers & $\log M_h$ & $c$ & $R_{\rm sig}$ & $\log
         M_{sub}$ & $f_{sub}$  \\ \\
  \hline\\
  M1 & [12.5-13.0) & BCG & $12.50^{+0.07}_{-0.07}$ & $11.35^{+2.41}_{-2.57}$ & $0.04^{+0.03}_{-0.02}$ & $10.1^{+1.39}_{-1.35}$ & $0.49^{+0.34}_{-0.33}$ \\
  M2 & [13.0-13.5) &BCG & $13.11^{+0.14}_{-0.12}$ & $9.19^{+3.45}_{-2.91}$ & $0.05^{+0.07}_{-0.03}$ & $11.45^{+1.07}_{-2.27}$ & $0.57^{+0.30}_{-0.35}$ \\
  M3 & [13.5-14.0) & BCG & $13.62^{+0.12}_{-0.10}$ & $9.19^{+3.45}_{-2.92}$ & $0.05^{+0.08}_{-0.03}$ & $11.44^{+1.07}_{-2.26}$ & $0.56^{+0.29}_{-0.35}$  \\
  M4 & [14.0-above) & BCG & $14.10^{+0.16}_{-0.10}$ &$9.64^{+3.09}_{-2.64}$ & $0.08^{+0.13}_{-0.06}$ & $13.31^{+0.33}_{-1.73}$ & $0.72^{+0.19}_{-0.32}$  \\ \\
  \hline
\end{tabular}
\end{center}
\end{table*}

\begin{table*}
\begin{center}
  \caption{\label{tab:tbl-4} Posterior of parameters we fitted to our
    X-ray luminosity high and low subsamples}
\begin{tabular}{llccccc}
\hline \hline \\
Mass bin & Centers+X ray & $\log M_h$ & $c$ & $R_{\rm sig}$ & $\log
             M_{sub}$ & $f_{sub}$  \\ \\
\hline \\
\multirow{4}{*}{13.0-13.5} &BCG+high  & $13.20^{+0.08}_{-0.08}$ & $11.11^{+2.79}_{-3.61}$ & $0.18^{+0.07}_{-0.09}$ & 
$12.06^{+0.52}_{-1.95}$ & $0.61^{+0.27}_{-0.31}$\\
  & BCG+low   & $12.89^{+0.13}_{-0.13}$ & $9.75^{+3.79}_{-3.60}$ & $0.11^{+0.17}_{-0.08}$ & $11.81^{+0.58}_{-2.25}$&  $0.63^{+0.25}_{-0.35}$  \\
  & X ray+high   & $13.19^{+0.09}_{-0.08}$ & $10.86^{+2.93}_{-3.63}$ & $0.17^{+0.07}_{-0.10}$ &$11.82^{+0.68}_{-2.25}$&  $0.58^{+0.29}_{-0.34}$ \\
  & X ray+low    & $12.89^{+0.13}_{-0.13}$ & $9.72^{+3.75}_{-3.52}$ & $0.11^{+0.19}_{-0.08}$ & $11.88^{+0.53}_{-2.15}$& $0.51^{+0.32}_{-0.34}$  \\\\
\hline \\
\multirow{4}{*}{13.5-14.0} & BCG+high   & $13.59^{+0.16}_{-0.10}$ & $6.80^{+2.51}_{-1.51}$ & $0.06^{+0.08}_{-0.04}$ & 
$11.11^{+1.37}_{-2.09}$ & $0.49^{+0.34}_{-0.33}$ \\
  & BCG+low   & $13.33^{+0.12}_{-0.09}$ & $11.28^{+2.37}_{-2.43}$ & $0.04^{+0.05}_{-0.03}$ & $10.96^{+1.52}_{-2.03}$& $0.52^{+0.32}_{-0.34}$  \\
  & X ray+high  & $13.58^{+0.12}_{-0.10}$ & $6.87^{+3.01}_{-1.61}$ & $0.09^{+0.07}_{-0.05}$ & $10.70^{+1.52}_{-1.82}$& $ 0.48^{+0.34}_{-0.35}$  \\
  & X ray+low   & $13.34^{+0.12}_{-0.10}$ & $11.29^{+2.39}_{-2.54}$ & $0.04^{+0.06}_{-0.03}$ &$10.99^{+1.49}_{-2.01}$& $ 0.53^{+0.32}_{-0.35}$  \\\\
\hline\\
\multirow{4}{*}{14.0-above} & BCG+high  & $14.05^{+0.17}_{-0.11}$ & $7.24^{+1.71}_{-1.33}$ & $0.04^{+0.05}_{-0.02}$& 
$11.86^{+1.30}_{-2.58}$ & $0.51^{+0.32}_{-0.32}$ \\
  & BCG+low   & $13.81^{+0.18}_{-0.16}$ & $10.36^{+3.07}_{-3.72}$ & $0.05^{+0.15}_{-0.03}$ & $12.67^{+0.68}_{-2.80}$& $0.69^{+0.23}_{-0.37}$   \\
  & X ray+high   & $14.01^{+0.14}_{-0.08}$ & $7.17^{+4.85}_{-2.29}$ & $0.23^{+0.12}_{-0.18}$ & $12.62^{+0.67}_{-2.42}$& $0.56^{+0.29}_{-0.30}$ \\
  & X ray+low    & $13.82^{+0.18}_{-0.16}$ & $9.88^{+3.37}_{-3.72}$ & $0.07^{+0.17}_{-0.05}$ &$12.83^{+0.55}_{-2.69}$& $ 0.70^{+0.29}_{-0.30}$  \\\\
\hline
\end{tabular}
\end{center}
\end{table*}

\subsection{Constrain halo properties using MCMC}
\label{results}

In this subsection, we present our constraining of the halo properties
using the above measured ESDs.  There are five free parameters in our
fitting process: host halo mass ($\log M_h$), concentration ($c$), off
center distance ($R_{\rm sig}$), subhalo mass ($\log M_{sub}$), and
subhalo fraction ($f_{sub}$).  We apply emcee
(http://dan.iel.fm/emcee/current/) to run a Monte-Carlo Markov Chain
(hereafter MCMC) to explore the likelihood function in the
multi-dimensional parameter space\footnote{Emcee is an MIT licensed
  pure-Python implementation of Goodman \& Weare’s Affine Invariant
  Markov chain Monte Carlo Ensemble sampler.}.

We assume a Gaussian likelihood function using convariance matrix
built from bootstrap sampling,
\begin{equation}
ln\mathcal{L}(\mathbf{X}|\mathbf{\Theta})=
-0.5((\mathbf{X}-\mathbf{D_{model}})^TC^{-1}(\mathbf{X}-\mathbf{D_{model}}))
\end{equation}
where $\mathbf{X}$ is the ESD data vector, $\mathbf{D_{model}}$ is the
model and $C^{-1}$ is the inverse of the covariance
matrix. $\mathbf{\Theta}$ denotes the parameters in the model.

In order to minimize the prior influence, we use broad flat priors for
all five parameters. We set halo mass range for each fitting mass bin
to be $[12.0, 16.0]$, concentration range $[1.0, 20.0]$, $R_{\rm sig}$
range $[0.0, 1.0]$ and $f_{sub}$ range $[0.0,1.0]$. The only physical
assumption on prior is that $10.0< \log M_{sub} < \log M_{h}-0.3$.
The lower limit for $\log M_{sub}$ is set since below which its ESD
signal can be neglected. As we have seen in Fig \ref{fig:massbins}
that the BCGs are the best tracers of the halo centers, we focus only
this set of ESD measurements.

Fig. \ref{fig:posterior} is an example of marginalized posterior
distributions of the five parameters for ESDs in the M1 sample after
MCMC process.  Within these five free parameters, we see only the host
halo mass $\log M_h$ and the off-center distance $R_{\rm sig}$ can be
well constrained.  While concentration $c$ is quite strongly correlated
with the off-center distance $R_{\rm sig}$.  The constraints on
the subhalo fraction and subhalo mass is very weak.

We show in Fig. \ref{fig:fitted} the best fitting results for groups
separated in four different halo mass bins as well as 
the ESD contributions for different mass components.  In the mass and
radius ranges we consider, the subhalo and 2-halo term contributions
are quite negligible. This is also the reason that we can't make tight
constraints on the subhalo fraction and subhalo mass.  The
contribution of stellar mass of the BCG is only important at very
small scales. Thus the observed ESDs in this study can mainly provide
us the constraints on the host halo properties.

Table \ref{tab:tbl-3} lists the best fitted parameters for groups in
different mass bins.  As we have seen from the likelihood distribution
of parameters in Fig. \ref{fig:posterior}, we can have fairly good
constrain on the halo mass of the group samples. However, if we
compare the halo masses obtained from the ESDs with those obtained
from the group catalog (c.f. Table 1), they are roughly underestimated
by 0.1$\sim$0.2 dex. We will discuss this discrepancy in the following
section.  The overall off-center distances for our BCGs are quite
small, assuring that BCGs are indeed good tracers of halo centers. On
the other hand, the concentrations of the halos seem to be somewhat
larger than the theoretical predictions \citep[e.g.][]{Zhao2009}.
However, since the concentration $c$ and the off-center distance
$R_{\rm sig}$ are quite tightly correlated, if we adopt the lower
value of $R_{\rm sig}$, the concentration $c$ will drop significantly
as well.

Listed in Table \ref{tab:tbl-4} are the best fitted parameters for
groups that are separated into X-ray luminosity high and low
subsamples.  Although, due to the smaller number of lens systems, the
error for each data point for our X-ray subsamples are somewhat larger
and thus the constraints on the five parameters are somewhat weaker,
we do see a prominent feature that the halo masses of high X-ray
luminosity subsamples are higher than their low luminosity
counterparts in all mass bins.  The difference is at 0.2$\sim$0.3 dex
which means that the high X-ray luminosity subsamples are nearly by a
factor of two more massive than their low X-ray luminosity
counterparts. 

% We do not further explore the $L_X$ and halo mass relation due to
% the fact that $L_X$ from \cite{Wang2014} catalog, mostly estimated
% using fixed $L_X-M_h$ relation as rough estimation based on
% \cite{Mantz2010}, except for 847 groups with X-ray detection larger
% than 3 sigma.

\begin{figure*}
\centering
\includegraphics[width=7.5cm,height=7.5cm]{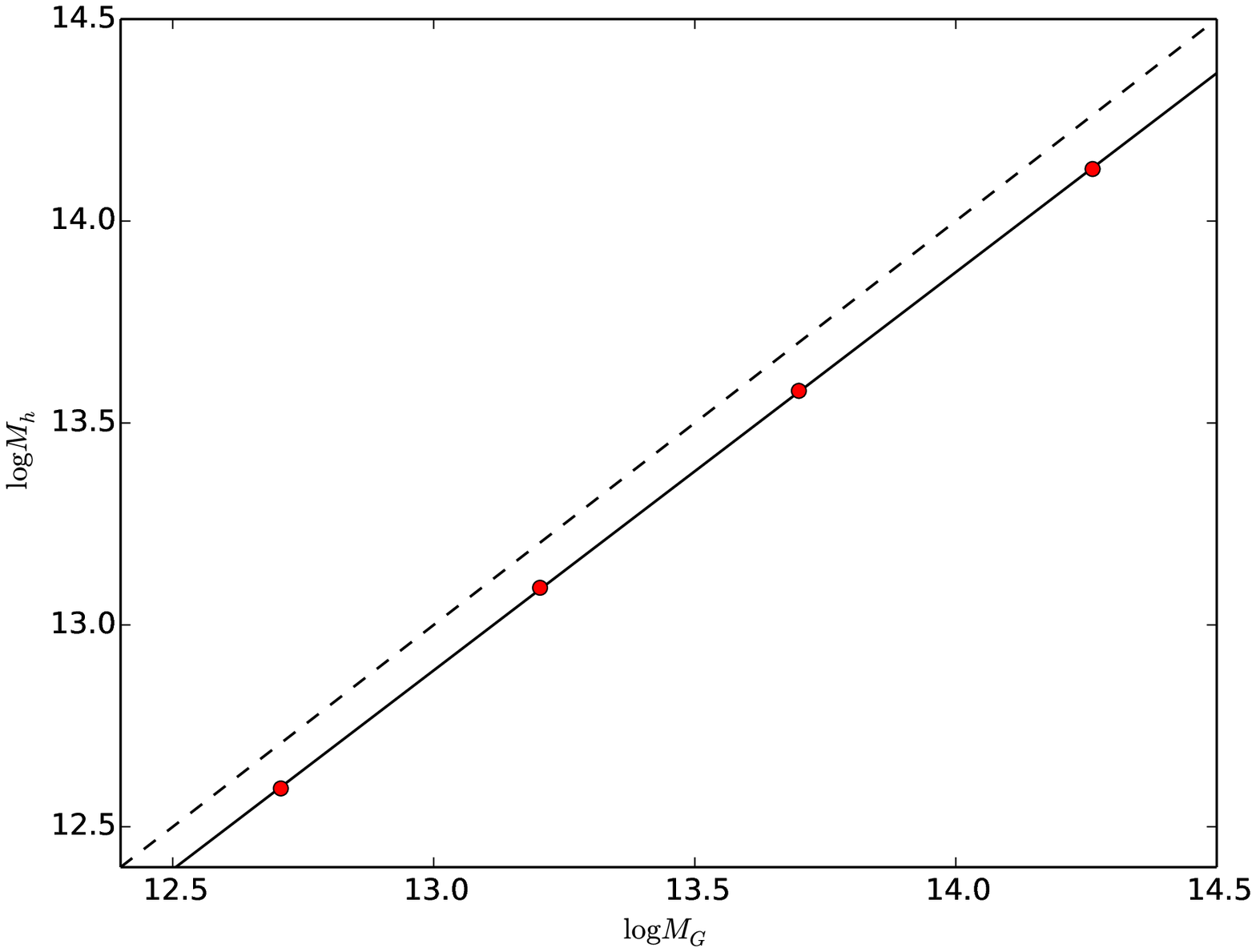}
\includegraphics[width=7.5cm,height=7.5cm]{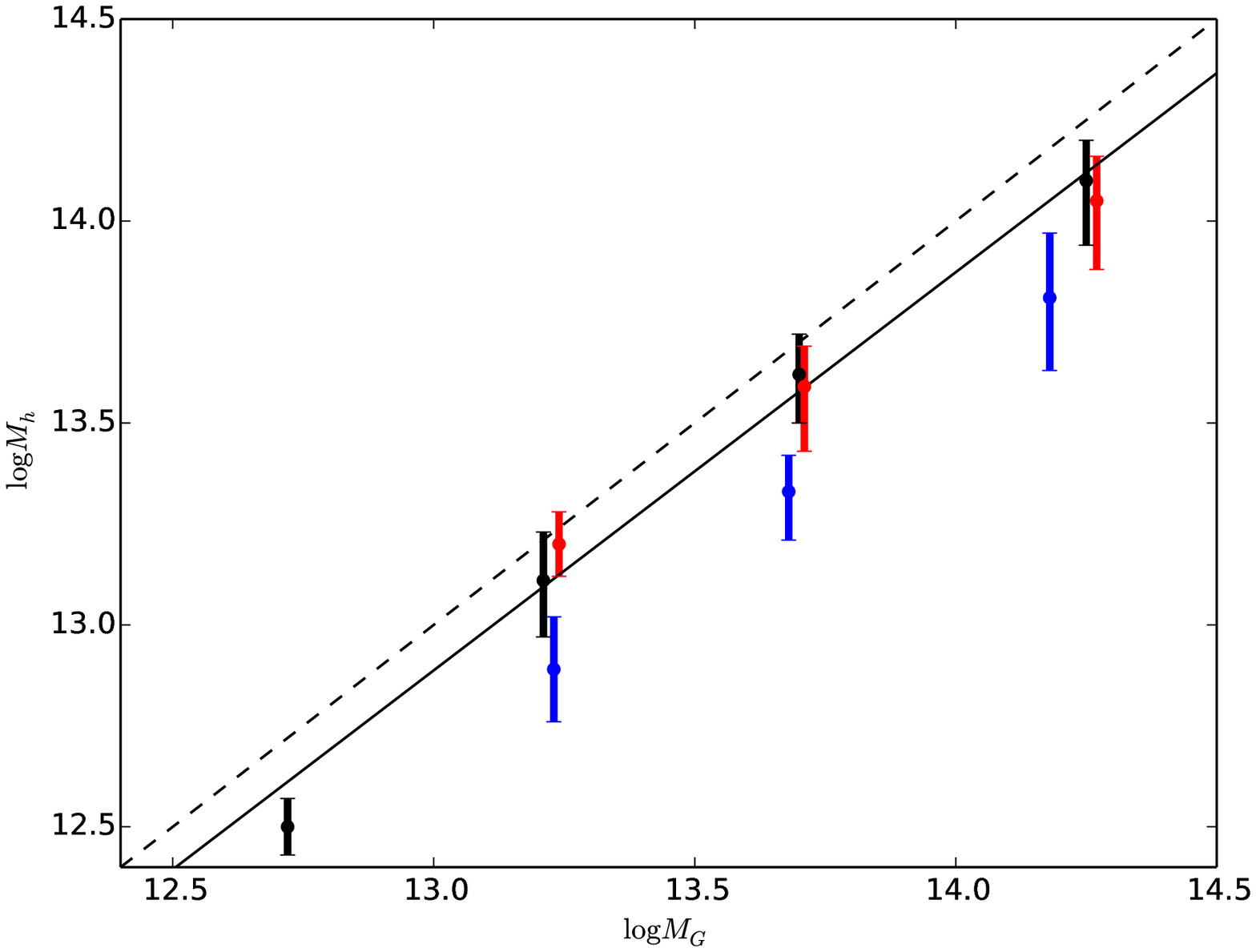}
\caption{Left panel: the average group mass v.s. the true halo mass
  for those groups whose mass estimation has an uncertainty at about
  0.3dex. The solid line, with $\log M_h= 0.9858*\log M_G- 0.07168$,
  shows the best fit results.  While the dashed line is the one-to-one
  relation.  Right panel: the average group mass estimated using
  luminosity ranking in Y07 v.s. the halo mass estimated using the
  galaxy-galaxy lensing signals. The black dots are the results from
  all the groups binned in 4 group mass bins. The red and blue ones
  are for the high X ray subsamples and low X ray subsamples. The
  solid line is the same as the one in the left panel. }
  \label{fig:massbias}
\end{figure*}

\section{Eddington bias of the halo mass estimation}
\label{sec_bias}

Recent studies have shown that the combination of galaxy-galaxy
lensing and clustering of galaxies can be used to constrain cosmology
\citep[e.g.][]{Bosch2013, More2013, Cacciato2013, Leauthaud2017}, the
halo masses estimated for groups using galaxy-galaxy lensing signals
and hence the halo mass function can also be used to constrain the
cosmological parameters. However, before doing so, one needs to make a
careful study of the systematics between the galaxy-galaxy lensing
measurement and modelling. Note that the group masses estimated from
the ranking of characteristic group luminosity or stellar mass have a
typical uncertainty at about 0.3 dex. Thus the halos/groups binned in
terms of group mass may induce an Eddington bias in the galaxy-galaxy
lensing halo mass estimation.

Here we make use of a high resolution N-body simulation to help us
understanding the systematics of galaxy-galaxy lensing modelling.  Our
simulation includes $3072^{3}$ dark matter particles in a periodic box
of $500 \mpch$ on a side \citep{Li2016}, which was carried out at the
Center for High Performance Computing at Shanghai Jiao Tong
University. It was run with {\tt L-GADGET}, a memory-optimized version
of {\tt GADGET2} \citep{Spr2005}. The cosmological parameters adopted
by this simulation are consistent with the WMAP9 results
\citep{Hin2013} (which are very similar to WMAP7 results as well), and
each particle has a mass of $3.4 \times 10^{8}\msunh$.  Dark matter
halos are identified using the standard friends-of-friends algorithm
with a linking length that is 0.2 times the mean inter particle
separation.  The mass of halos, $M_h$, is simply defined as the sum of
the masses of all the particles in the halos, and we remove halos with
less than 20 particles.

Using the true halos in the simulation, we mimic the halo mass
estimation uncertainty in the groups as follows. First, we add to each
halo mass a Gaussian scatter with $\sigma =0.3$ in log space. Next, we
rank all the resulting halos and match them with the halo mass
function model prediction \citep{Tinker2008} assuming a WMAP7
cosmology to assign each halo a new mass.  Thus assigned halo masses
are referred to as the group masses $M_G$.  Following the same mass
selection criteria used for our galaxy-galaxy lensing studies, we
separate the groups (halos) into four samples. From these four
samples, we estimated both the average group mass $M_G$ and the true
halo mass $M_h$.  The resulting group v.s. true halo mass relation is
shown in the left panel of Fig. \ref{fig:massbias}.  Obviously, the
data points are off from the diagonal dotted line at about 0.1 dex
level, which illustrates a bias between these two halo masses.  We use
a solid line to fit the data points, which can be used to roughly
account for the Eddington bias in our weak galaxy-galaxy lensing
studies.

Shown in the right panel of Fig. \ref{fig:massbias} is the average
group mass v.s. the true halo mass estimated from the galaxy-galaxy
lensing signals in our study.  For all the group sample, after taking
into account the Eddington bias, i.e. comparing to the solid line, the
data in the three massive group bins agree very well. Indicating the
WMAP7 cosmology adopt in this study is quite consistent with the
lensing mass constraints.  On the other hand, according to the result
for the lowest mass bin, we still see that the halo mass estimated
from the galaxy-galaxy lensing signals is about 0.1 dex lower.
Similar trends were also reported in a recent study by
\cite{Leauthaud2017}, where they found that the lensing signals
predicted from clustering are 20\%-40\% larger than the true
measurements. It still remains unclear to us what is the main cause of
this lensing deficiency around relatively low mass halos.

While for the groups that are separated into X-ray luminosity high and
low subsamples, we do see that the average halo masses of X-ray
luminosity low subsamples are systematically smaller.  Thus a
combination of X-ray luminosity and optical total group luminosity
will be useful to better constrain the individual group/cluster mass. 

Finally, we caution that when using galaxy-galaxy lensing signals
around lens systems to constrain cosmological parameters in future
larger surveys, it would be important to take into account the
Eddington bias as demonstrated here.

\section{Summary and Conclusion}
\label{summary}

We measure the galaxy-galaxy lensing signals around group samples in
different mass bins using the source galaxy shape measurements
obtained by \citet{Luo2017}, where the group masses are provided by
\cite{Yang2007} using the ranking of characteristic group luminosity.
We also divide the groups with X-ray luminosities obtained by
\citet{Wang2014} into X-ray luminosity high and low subsamples, and
measured their galaxy-galaxy lensing signals separately.

We then model the galaxy-galaxy signals by considering contributions
from an off-centered NFW profile, sub halo, stellar mass and 2-halo
term using five free parameters: halo mass $\log M_h$, concentration
$c$, off-center distance $R_{\rm sig}$, subhalo fraction $f_{sub}$ and
the subhalo mass $\log M_{sub}$. We then run MCMC to constrain the
five free parameters by assigning them with flat priors. From the
lensing signals we measured from the SDSS DR7 observation, we are able
to provide relatively good constraints on the halo properties, however not
the subhalo properties. Below we summarize the main findings of this
work.

\begin{itemize}
\item By checking the galaxy-galaxy lensing signals around four kinds
  of halo center tracers: BCG, luminosity weighted center, number
  weighted center and X-ray peak position, we find that BCG is the best
  halo center tracer.

\item The off-center effect $R_{\rm sig}$ for BCG is roughly at
  $0.05\mpch$, from $0.04\mpch$ for the lowest mass bin group sample
  to $0.09\mpch$ for the most massive group sample.

\item After taking into account the Eddington biasThe, halo masses
  estimated from galaxy-galaxy lensing signals are consistent with the
  group masses obtained using abundance matching assuming WMAP7
  cosmology in the three massive samples.  This consistency implies
  that the WMAP7 cosmology is favored by the lensing signals measured
  in this study.

\item If separated into X-ray luminosity high and low subsamples, X
  ray luminosity low subsamples have overall lower halo masses
  compared to their X ray luminosity high counterparts. This X-ray
  luminosity segregation in halo mass indicates that we can combine X
  ray luminosities and optical luminosities of groups to better
  constrain their individual masses.
\end{itemize}

Finally, as the galaxy-galaxy lensing can provide us an independent
measurement of the halo mass for clusters and groups, for larger and
deeper surveys, we can use the resulting halo mass functions to
constrain cosmological parameters.

\acknowledgements

This work is supported by the 973 Program (No. 2015CB857002,
2015CB857001), national science foundation of China (Nos. 11233005,
11503064, 11621303), Chinese Scholarship Council (201504910477) and
Shanghai Natural Science Foundation, Grant No.  15ZR1446700. JZ  
is  supported  by  the  NSFC  grants  (11673016, 11433001, 11621303) and 
the National Key Basic Research Program
of China (2015CB857001). L.P.F. acknowledges the support from NSFC grant 11333001 and
11673018, STCSM grants 13JC1404400 and 16R1424800, SHNU grant DYL201603. We also
thank the support of the Key Laboratory for Particle Physics,
Astrophysics and Cosmology, Ministry of Education.

\end{document}